\documentclass[aps,prb,twocolumn,showpacs,amsmath,amssymb,superscriptaddress,groupedaddress]{revtex4-1}
\usepackage{graphics}
\usepackage{epsfig}
\usepackage[colorlinks=true,citecolor=blue,linkcolor=red,linktocpage=true,pagebackref=false]{hyperref}
\usepackage{soul} 
\usepackage[usenames, dvipsnames]{color}
\usepackage{braket}

\begin{document}
   \title{
First-order correlation function of the stream of single-electron wave-packets
}
\author{Michael Moskalets}
\email{michael.moskalets@gmail.com}
\affiliation{Department of Metal and Semiconductor Physics, NTU ``Kharkiv Polytechnic Institute", 61002 Kharkiv, Ukraine}

\date\today
\begin{abstract}
The first-order correlation function, which is accessible experimentally, contains all essential information about the state of the system of non-interacting electrons.
Here I discuss how this function can be used to answer the question whether the state of a periodic stream of single-electron wave-packets is a multi-particle state or it is the product of single-particle states. 
In the latter case the correlation function is expected to be factorizable while in the former case it is not. 
As an example I consider a train of Lorentzian in shape single-electron excitations, levitons. 
I demonstrate that the correlation function in time domain is factorizable or not depending on whether the wave-packets are separated or overlapping. 
In contrast, the correlation function in energy domain is always factorizable and thus cannot be used to distinguish single- and multi-particle states. 
\end{abstract}
\pacs{73.23.-b, 72.10.-d, 73.63.-b}
\maketitle

\section{Introduction} 

Recently Jullien {\it et al.}\cite{Jullien:2014ii} have reported a 
 measurement of a wave function of a single electron traveling in a ballistic conductor. 
This groundbreaking experiment opens the route to the control and characterization of single- and few- electron states injected on demand into solid state circuits.\cite{Blumenthal:2007ho,Feve:2007jx,Fujiwara:2008gt,Kaestner:2008hp,Bocquillon:2013dp,Fricke:2013cc,Dubois:2013dv,Fletcher:2013kt,Tettamanzi:2014gx,Ubbelohde:2014vx} 

\begin{figure}[t]
\includegraphics[width=80mm]{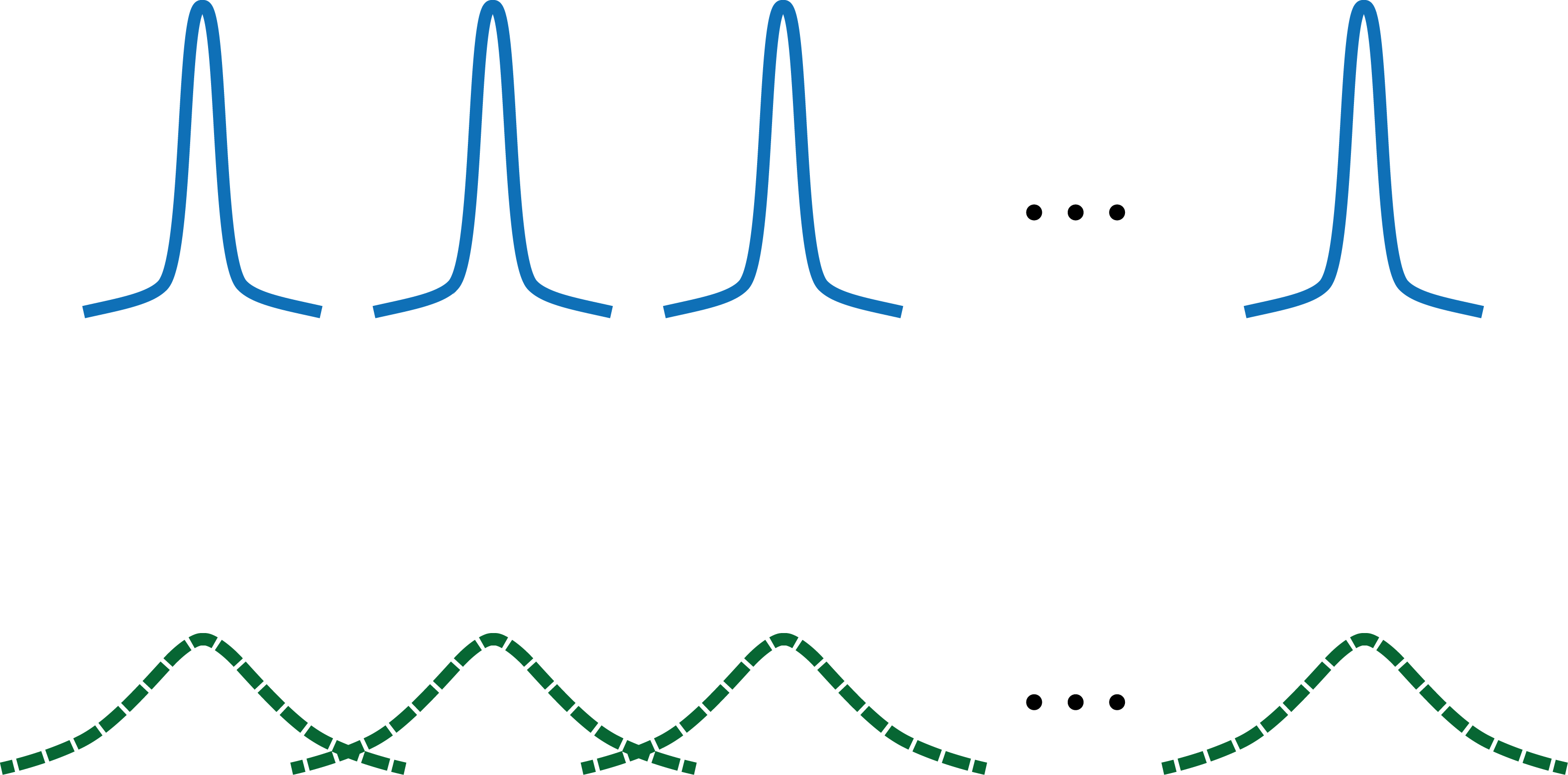}
\caption{(Color online) The periodic stream of single-electron wave-packets: If the width of an individual wave-packet is smaller than the period, i.e., the distance between subsequent wave-packets,  then all electrons are distinguishable by the period they occupy and no a multi-particle state is formed. In particular during each period only one electron can be detected. If in contrast the width of a wave-packet is comparable with the period then more then one electron can be detected during a period, though the mean number of detected particles remains the same. Due to the overlap between different wave-packets the multi-particle state is formed.}
\label{fig1}
\end{figure}

The experiment of Ref.~\onlinecite{Jullien:2014ii} was done with a train of electrons, each prepared in a well-controlled single-electron state called a leviton. 
Such a train is also appropriate to address another problem important for a few-electron state engineering. 
Namely, it is important to demonstrate directly that a multi-particle state is not  formed in the stream of levitons. 
Such a multi-particle state should be necessarily formed if the distance between the  levitons would become comparable with the width of a leviton, see Fig.~\ref{fig1}. 
So, merely by varying the rate at which the levitons are generated, one can tune the state of a stream from a multi-particle state to (the product of) single-particle states. 

In quantum optics\cite{Walls:2008uy} the single- and multi-photon states are distinguished with the help of the intensity-intensity correlation measurement\cite{Kuhn:2002ds,Bozyigit:2010bw}: 
The photon flux is divided by a beam splitter and the  resulting fluxes are directed to two single-photon detectors. The absence of coincident detections proves that a multi-photon state is not present in the flux.  
This technique is customarily used to characterize single-photon sources.\cite{Lounis:2005ex} 

In quantum coherent electronics the analogue of an optic beam splitter is a quantum point contact,\cite{VanWees:1988vf,Wharam:1988vi} which divides an electric current into two beams, the reflected and the transmitted one. 
Thus the current cross-correlation measurement is expected to be an analogue of a coincidence detection measurement in optics. 
However this is not the case. 
The reason is that there are no efficient single-electron detectors available at the moment. 
Therefore, what is measured in electronics is a current noise, the product of  fluctuations of reflected and transmitted currents averaged over a long time. 
Apart of the thermal fluctuations vanishing at low temperatures, the noise contains a part, known as the shot noise, which relies on the quantization of charge and exists due to the fact that an electron can be either transmitted or reflected but not both.\cite{Blanter:2000wi}
In the absence of electron-electron correlations the shot noise is proportional to the  mean number of particles in a stream per unit time,\cite{Lee:1995tv,Levitov:1996ie,Ivanov:1997wz,Keeling:2006hq,Olkhovskaya:2008en} since each particle is scattered independently. 
This fact was used in Refs.~\onlinecite{Bocquillon:2012if,Dubois:2013dv} to count directly the number of particles emitted in each cycle. 
Another approach to characterize a single-electron emission regime is to measure a finite frequency noise, which also provides information on a short-time dynamics of the source.\cite{Mahe:2010cp,Albert:2010co,Parmentier:2012ed,Moskalets:2013ed}

If the distance between particles within a periodic stream would decrease such that different particles start to overlap, see Fig.~\ref{fig1}, the mean number of particles  per period would not change while a multi-particle state would emerge. 
This results in fluctuations of the number of particles measured during a given period.   
The shot noise, however, cannot distinguish single- and multi-particle states as long as the  mean number of particles per period is the same.  

The aim of this work is to demonstrate that another measurable, namely the first-order correlation function, can be used to make the necessary distinction. 
In Refs.~\onlinecite{Haack:2011em,Haack:2013ch} it was shown that the correlation function in time domain is  directly accessible via a time resolved interference current, which is measured at the exit of an electronic Mach-Zehnder interferometer\cite{Ji:2003ck}. 
Another approach to access a correlation function but in energy domain is a single-electron quantum tomography.\cite{Grenier:2011js,Grenier:2011dv,Ferraro:2013bt,Ferraro:2014ev} It relies on a zero-frequency noise measurement in the Hanbury Brown and Twiss set-up\cite{Brown:1958uu,Henny:1999tb,Oliver:1999ws}, where the source of interest is in one input channel while the other input channel is fed by a weak ac voltage. 
Such an electronic quantum tomography was realized in Ref.~\onlinecite{Jullien:2014ii}.

The paper is organized as follows: 
In Sec.~\ref{sec1} some basic properties of fermionic correlation functions are presented and the energy representation in the case of a periodic flux is introduced. 
In Sec.~\ref{lev} the correlation function of a periodic train of electron-like excitations, levitons, is analyzed in both time and energy domains. 
The conclusion is given in Sec.~\ref{concl}. 
Some details of calculations are presented in Appendices \ref{proof2} - \ref{ac}.

\section{Excess correlation function}
\label{sec1}

Let us consider an electronic one-dimensional chiral (or ballistic) waveguide connected to an electronic reservoir in equilibrium, a metallic contact at Fermi energy $ \mu$, which serves as a source of electrons. 
The first-order correlation function for spinless electrons in a waveguide is defined as follows, $\tilde {\cal G}^{(1)}\left(1;2  \right) =
\langle \hat\Psi^{\dag}(1) \hat\Psi(2) \rangle$, where $\hat\Psi (j) \equiv \hat\Psi\left(x_{j},t_{j} \right)$ is an electron field operator in the second quantization evaluated at point $x_{j}$ and time  $t_{j}$,  $j=1,2$. 
The quantum-statistical average $\langle \dots \rangle$ is taken over the equilibrium state of the electronic reservoir. 

If a voltage $V(t)$ is applied on the contact, then a non-equilibrium flux of particles is injected from a contact into a waveguide. 
The excess first-order correlation function is defined as the difference of correlation functions with the voltage on and off (the subscripts ``$V$'' and ``$0$'', respectively): $\tilde G^{(1)} = \tilde {\cal G}^{(1)}_{V} - \tilde {\cal G}^{(1)}_{0}$.\cite{Grenier:2011js,Grenier:2011dv} 
The correlation function $\tilde G^{(1)}$ characterizes a flux of particles injected from a driven  contact into a waveguide on the top of the Fermi sea with chemical potential $ \mu$, which plays a role of a fermionic vacuum. 
Importantly, the contribution of injected particles is completely separated from that of the underlying Fermi sea.\cite{Bocquillon:2013fp} 

At zero temperatures the disturbance of the vacuum due to the injected flux consists in the appearance of particles with energy larger than $ \mu$ and in the disappearance of some particles with energy less than $ \mu$.  
In the former case one can speak about injection of (quasi-)electrons, while in the latter case about injection of holes.  
Below I will focus on an electron injection. 
However the presence of holes can be also taken into account by adding a hole correlation function.\cite{Grenier:2011js,Grenier:2011dv,Bocquillon:2013fp} 

Here I am interested in the case when the amplitude of the voltage is relatively small, $eV \ll \mu$, and, therefore, the spectrum of injected electrons in a waveguide can be linearized close to the Fermi energy $  \mu$. 
In this case the excess correlation function depends on a combined coordinate, $t_{j} \equiv t_{j} - x_{j}/v_{ \mu}$ with $v _{ \mu}$ the Fermi velocity, rather than on $x_{j}$ and $t_{j}$ separately.

\subsection{A pure $N$-electron state}

The single-particle correlation function for a pure $N$-electron state (with possibly $N\to \infty$) can be represented as follows, see e.g. Ref.~\onlinecite{Grenier:2013gg}, 

\begin{eqnarray}
\tilde G ^{(1)}_{N}  (t_{1} ;t_{2} )
= \sum _{ k = 1} ^{N} \psi _{ k} ^{*}(t_{1} ) \psi _{ k} (t_{2} ) .
\label{g1}
\end{eqnarray}
\ \\
\noindent
where the mode wave functions $\psi _{ k}$ are orthonormal, 

\begin{eqnarray}
v_{ \mu} \int dt \psi _{ k _{1}} ^{*}(t) \psi _{ k _{2} } (t) =  \delta _{ k _{1} k _{2} } .
\label{psi-ortho}
\end{eqnarray}

However, the decomposition presented in Eq.~(\ref{g1}) is not always known a priori, especially if  the correlation function is determined experimentally. 
In this case the following property of the correlation function is useful to prove that an electronic state in question is a pure state,

\begin{eqnarray}
v_{ \mu} \int dt  \tilde G ^{(1)}  (t _{1} ;t ) \tilde G ^{(1)} (t  ;t _{2} ) =    \tilde G ^{(1)}  (t _{1} ;t _{2} ) .
\label{conv-1}
\end{eqnarray}

As it is pointed out in Ref.~\onlinecite{Beenakker:2005gh}, an oscillating potential acting on non-interacting electrons of the Fermi sea generates a pure state. 
The state emitted by a single-electron source without intrinsic dephasing\cite{Haack:2013ch,Iyoda:2014cf} is also expected to be pure.  

The higher-order (excess) correlation function is expressed in terms of the first-order correlation function via the Slater determinant, 

\begin{eqnarray}
 \tilde G ^{(M)}  (f _{1},f _{2},\dots,f _{M};i_{M},i _{M-1},\dots,i _{1}) = 
\nonumber \\
\label{cf-03} \\
= \det
\left(
\begin{array}{ccc}
  \tilde G ^{(1)}  (f _{1} ;i _{1} ) & \dots &  \tilde G ^{(1)}  (f _{1} ;i _{M} )    \\
  \dots & \ddots  & \dots  \\
  \tilde G ^{(1)}  (f _{M} ;i _{1} ) &  \dots &  \tilde G ^{(1)}  (f _{M} ;i _{M} )    
\end{array}
\right) ,
\nonumber 
\end{eqnarray}
\ \\
\noindent
where $f$ and $i$ stand for the final and initial coordinates, respectively. 
The derivation of the second-order excess correlation function is presented in Ref.~\onlinecite{Moskalets:2014ea}. 
The higher-order correlation functions can be calculated along the same lines.  

Remarkably, for a fermionic $N$-particle state the correlation functions of order higher than $N$ all nullify, (see Appendix \ref{proof2})

\begin{eqnarray}
\tilde G ^{(M)}_{N}  (f _{1},f _{2},\dots,f _{M};i_{M},i _{M-1},\dots,i _{1}) = 0 , \quad M > N . 
\nonumber \\
\label{pure-NM}
\end{eqnarray}
\ \\
\noindent 
Therefore, for a single-electron state $G^{(2)} \equiv 0$, which is a consequence of the fact that the first-order correlation function is factorizable, that is the sum in Eq.~(\ref{g1}) contains only one term. 
This property can be used to demonstrate a single-electron state.

\subsection{The energy representation} 

Sometimes the energy representation is more convenient than the time representation. 
Especially if both electrons and holes are present.\cite{Grenier:2011js,Grenier:2011dv,Bocquillon:2013fp}  

In general in the energy representation the two-time quantities  like $G^{(1)}\left( t_{1}; t_{2} \right)$  become dependent on two energies.
The energy and time representations are related via a corresponding Fourier transformation. 
However in the case of a system driven periodically with period ${\cal T} _{0} = 2 \pi / \Omega$, the difference of two energies is not arbitrary but can only be a multiple of the energy quantum $ \hbar \Omega$. 
It is convenient to account this property directly and denote one energy as $E$ and the other one as, say, $E_{m} = E + m \hbar \Omega$ with $m$ an integer.  
I introduce two correlation functions in the energy representation, $\tilde G ^{(1),in}$ and $\tilde G ^{(1),out}$, depending on which time $t_{2}$ or $t_{1}$, respectively, can be viewed as conjugate to $E$. 
The other time can be viewed is conjugate to $E_{m}$. 
The formal definitions are the following, 

\begin{subequations}
\label{pur0203}

\begin{eqnarray}
\tilde G ^{(1),in} (E_{m} ; E) 
&=&
\int _{ - \infty}^{ \infty } d \Delta t   e ^{- i E  \frac{ \Delta t }{ \hbar }} 
\nonumber \\
\label{pur-03-1} \\
&&
\times
\int _{ 0 }^{ {\cal T} _{0} } \frac{ d t _{1} }{ {\cal T} _{0} }
e ^{  -i m \Omega   t _{1}} 
\tilde G ^{(1)} _{} (t _{1}  ; t _{1} - \Delta t ) ,
\nonumber 
\end{eqnarray}

\begin{eqnarray}
\tilde G ^{(1),out} (E; E_{m}) 
&=&
\int _{ - \infty}^{ \infty } d \Delta t   e ^{- i E \frac{ \Delta t }{ \hbar }} 
\nonumber \\
\label{pur-02-1} \\
&&
\times
\int _{ 0 }^{ {\cal T} _{0} } \frac{ d t _{2} }{ {\cal T} _{0} }
e ^{  i m \Omega   t _{2}} 
\tilde G ^{(1)} _{} ( \Delta t + t _{2} ; t _{2} ) ,
\nonumber 
\end{eqnarray}
\ \\ \noindent
where $ \Delta t = t_{1} - t_{2}$. 
Note at $m = 0$ both functions above are the same, $\tilde G ^{(1),in} (E; E) = \tilde G ^{(1),out} (E; E)$.

It is convenient to parametrize $E$ by the Floquet (quasi-)energy, $0 < \epsilon < \hbar \Omega$,\cite{Platero:2004ep} and by the Floquet band number, $n$, as follows $E =  \mu + \epsilon + n \hbar \Omega \equiv \epsilon_{n} $.  
With this notation the inverse Fourier transformations can be written as follows,

\begin{eqnarray}
\tilde G ^{(1)} _{} (t _{1}  ; t _{2} )
&=&
e ^{ i \mu \frac{ \Delta t }{ \hbar }} 
\int _{0}^{ \hbar \Omega } \frac{d \epsilon }{h }  e ^{ i \epsilon \frac{ \Delta t }{ \hbar }} 
\sum\limits_{ n=-\infty}^{\infty}
e ^{ i n \Omega \Delta t}
\nonumber \\
\label{pur-03} \\
&&
\times
\sum\limits_{ m=-\infty}^{\infty}
e ^{ i m \Omega   t _{1}} 
\tilde G ^{(1),in} (\epsilon_{n+m} ; \epsilon_{n}) .
\nonumber 
\end{eqnarray}

\begin{eqnarray}
\tilde G ^{(1)} _{} ( t _{1}; t _{2} ) 
&=&
e ^{ i \mu \frac{ \Delta t }{ \hbar }} 
\int _{0}^{ \hbar \Omega } \frac{d \epsilon }{h }  e ^{ i \epsilon \frac{ \Delta t }{ \hbar }} 
\sum\limits_{ n=-\infty}^{\infty}
e ^{ i n \Omega \Delta t} 
\nonumber \\
\label{pur-02} \\
&&
\times
\sum\limits_{ m=-\infty}^{\infty}
e ^{ - i m \Omega   t _{2}} 
\tilde G ^{(1),out} (\epsilon_{n}; \epsilon_{n+m}) ,
\nonumber 
\end{eqnarray}
\end{subequations}

Since by its definition the correlation function satisfies the following relation, $ \tilde G ^{(1)}(t _{1}; t _{2}) = \left [ \tilde G ^{(1)}(t _{2}; t _{1})  \right] ^{*}$, one can get for its Fourier transforms the following, $\left [  \tilde G ^{(1),in} (\epsilon_{n}; \epsilon_{\nu}) \right] ^{*} = \tilde G ^{(1),out} (\epsilon_{\nu} ; \epsilon_{n})$.
The latter allows to represent equations entirely either in terms of $\tilde G ^{(1),in}$ or in terms of $\tilde G ^{(1),out}$.

The use of two different Fourier transforms is especially convenient to calculate convolutions, like in the purity condition, Eq.~(\ref{conv-1}).   
Applying Fourier transformations presented in Eqs.~(\ref{pur-03-1}) and (\ref{pur-02-1}) to Eq.~(\ref{conv-1}) we arrive at the following relations specific for a pure electronic state,

\begin{eqnarray}
\tilde G ^{(1),in} (\epsilon_{n} ; \epsilon_{\nu}) 
&=& 
\tilde G ^{(1),out} (\epsilon_{n} ; \epsilon_{\nu}), 
\nonumber \\
\label{pur-06-new} \\
\tilde G ^{(1),in} (\epsilon_{n} ; \epsilon_{\nu}) 
&=&
v_{ \mu}
\sum\limits_{ m=-\infty}^{\infty}
\tilde G ^{(1),in} (\epsilon_{n} ; \epsilon_{m} ) 
\tilde G ^{(1),in} ( \epsilon_{m}; \epsilon_{\nu}  ) .
\nonumber 
\end{eqnarray}
\ \\ \noindent
The details of calculations are given in Appendix  \ref{Fourie-pure}.

\subsection{Correlation function for adiabatically emitted electrons}

The excess correlation function can be expressed in terms of the Floquet scattering amplitude\cite{Moskalets:2011cw} of a periodically driven source, which depends on two energies. 
However in the limit of an adiabatic (slow) drive, the source can be  characterized by the frozen scattering amplitude, $S(E,t)$, which depends on one energy and time. 
The dependence on energy $E$ is the same as for the scattering amplitude of a stationary source. 
The dependence on time $t$ stems from the fact that some parameters of a source (and hence of a stationary scattering matrix) are affected by the drive.  
Since the drive is periodic in time, the scattering amplitude is also periodic in time, $S(E,t) = S(E,t + {\cal T} _{0})$. 
At zero temperature what matters is the scattering amplitude of a source  evaluated for electrons with Fermi energy, $ E=\mu$. 
I denote such an amplitude as $S(t) \equiv S( \mu,t)$.  

In the case of a quantum-dot based source, for instance, a mesoscopic capacitor side-attached to a waveguide,\cite{Buttiker:1993wh,Gabelli:2006eg,Feve:2007jx} the regime of a drive is adiabatic if the dwell time (the time an electron spends within the source) is small compared to the time extent of an emitted wave-packet.\cite{Splettstoesser:2008gc,Moskalets:2013dl} 
If an electron stream is generated by applying an ac voltage $V(t)$ to a metallic contact as in Refs.~\onlinecite{Dubois:2013dv,Jullien:2014ii}, then the regime is adiabatic  if $ \hbar \Omega, eV \ll \mu$. 
The corresponding scattering amplitude is the following, $S(t) = e ^{ - i \frac{e }{ \hbar } \int ^{t} d t^{\prime} V(t^{\prime}) }$.\cite{Jauho:1994cg,Pedersen:1998uc,Moskalets:2008ii} 

At zero temperature and in the adiabatic driving regime, the excess first-order correlation function is expressed in terms of a scattering amplitude of the source $S(t)$ as follows, \cite{Haack:2011em,Haack:2013ch}

\begin{eqnarray}
\tilde G^{(1)}(t _{1} ; t _{2}) 
&=&
\frac{1 }{ v _{ \mu} }
e ^{ i \left(  t _{1} - t _{2} \right) \frac{ \mu }{ \hbar } } 
  G ^{(1)}(t _{1} ; t _{2}) 
,
\nonumber \\
\label{cf-01} \\
 G ^{(1)}(t _{1} ; t _{2}) 
&=&
\frac{ i  }{2 \pi  }
\frac{ 1 - S ^{*} (t _{1}) S( t _{2} ) }{ t _{1} - t _{2} } .
\nonumber 
\end{eqnarray}
\ \\
\noindent
The phase factor, $e ^{ i \left(  t _{1} - t _{2} \right) \frac{ \mu }{ \hbar } } $, and the constant factor $1/v _{ \mu}$ are not important for further discussion, therefore, I will work below with the envelope correlation function $ G ^{(1)}$ only.  
Note the absence of a factor $v_{ \mu}$ in Eqs.~(\ref{conv-1}), (\ref{pur-06-new}) written in terms of $G ^{(1)}$ not $\tilde G ^{(1)}$. 
In addition in Eqs.~(\ref{pur0203}) written in terms of $G ^{(1)}$ we have to put $ \mu = 0$. 

The correlation function $ G ^{(1)}(t _{1} ; t _{2})$ is not periodic in its arguments. 
However, since the scattering amplitude $S(t)$ is periodic, then, as it follows from Eq.~(\ref{cf-01}), the reduced correlation function, 

\begin{eqnarray}
g\left( t_{1} ; t_{2} \right) = \left( t_{1} - t_{2} \right) G^{(1)}\left( t_{1}; t_{2} \right) ,
\label{reduce}
\end{eqnarray}
\ \\
\noindent
is periodic in each of its arguments. 
Therefore, it is sufficient to measure $G^{(1)}$ over one period in each of its arguments and then to use the periodicity of $g\left( t_{1} ; t_{2} \right)$ to extend $G^{(1)}$ to larger arguments. 

Moreover, as it follows from Eq.~(\ref{cf-01}), only the diagonal part of the  correlation function contains all essential information. 
To show this let us take the limit $t _{1} \to t _{2}$ in Eq.~(\ref{cf-01}) and find  $G ^{(1)} \left( t;t \right) = (- i/ 2 \pi) S(t)  \partial S ^{*}(t) /\partial t $.
Since the scattering amplitude is unitary ($S ^{*}S = 1$), we can represent it as $S(t) = e ^{ i \phi(t) }$ and obtain $ \phi(t) = - 2 \pi \int _{}^{t } dt^{\prime} G ^{(1)}(t^{\prime};t^{\prime})$. 
Finally we rewrite Eq.~(\ref{cf-01}) in terms of the correlation function only,

\begin{eqnarray}
 G ^{(1)}(t _{1} ; t _{2}) 
&=&
\frac{ i  }{2 \pi  }
\frac{ 1 - e ^{ i 2 \pi \int _{ t _{2} }^{ t _{1} } dt^{\prime}  G ^{(1)}( t^{\prime};t^{\prime}) } }{ t _{1} - t _{2} } .
\label{cf-05}
\end{eqnarray}
\ \\
\noindent 
This relation, valid for a single-channel case, greatly simplifies an experimental set-up necessary to find a correlation function. 
Indeed, a time resolved measurement of the current generated by the ac source, $I(t) = (-ie/2 \pi) S \partial S ^{*} /\partial t $,\cite{Buttiker:1994vl,Avron:2000de} not an interferometric current,\cite{Haack:2011em,Haack:2013ch} is sufficient to access $G ^{(1)}(t _{1} ; t _{2})$ since $ G ^{(1)}(t ; t ) = I(t)/e$. 
If nevertheless $ G ^{(1)}(t _{1} ; t _{2})$ as a function of both its arguments is experimentally available, then Eq.~(\ref{cf-05}) can be used to ensure that the driven system is in the adiabatic regime.
For non-adiabatically driven systems the correlation function $\tilde G^{(1)}$ is related to the Floquet scattering matrix rather than to the frozen scattering amplitude of the source.\cite{Haack:2013ch} 
As a result,  equation (\ref{cf-05}) is not expected to hold.

\section{A periodic stream  of overlapping levitons}
\label{lev}

In Refs.~\onlinecite{Dubois:2013dv,Jullien:2014ii} the train of wave-packets  carrying each an elementary charge was generated by applying a periodic sequence of the Lorentzian voltage pulses with integer flux,

\begin{eqnarray}
eV_{st}(t) = \sum _{n=-\infty} ^{\infty} \frac{ 2 \hbar \Gamma _{ \tau} }{\left( t - n {\cal T} _{0}  \right)^2 + \Gamma _{ \tau}^2} .
\label{Vlev}
\end{eqnarray}
\ \\
\noindent 
to a metallic contact, as it was  suggested in Refs.~\onlinecite{Levitov:1996ie,Ivanov:1997wz}. 
These elementary excitations were named levitons. 
In the equation above  $\Gamma _{ \tau}$ is the half-duration of a single pulse and ${\cal T} _{0} = 2 \pi/ \Omega$ is a period, i.e., the time delay between two subsequent pulses. 
The subscript ``$st$'' is used to differentiate quantities related to a stream. 
For such a potential the corresponding scattering amplitude $S_{st}(t)$ reads,\cite{Dubois:2013fs}

\begin{eqnarray}
S _{st}(t) &=& \prod _{k = - \infty} ^{ \infty} S_{L}\left( t + k {\cal T} _{0} \right) 
= \frac{ \sin\left( \pi \frac{ t  + i \Gamma  _{ \tau} }{ {\cal T} _{0} } \right) }{ \sin \left(\pi \frac{ t  - i \Gamma  _{ \tau} }{ {\cal T} _{0} } \right) } ,
\label{st-01}
\end{eqnarray}
\ \\
\noindent
where $S_{L}(t) = \left( t + i  \Gamma _{\tau}  \right) / \left(  t  - i  \Gamma _{\tau}  \right) $ is a scattering amplitude corresponding to a single voltage pulse, which excites a single leviton. 
Using the equation above in Eq.~(\ref{cf-01}) we obtain a correlation function,

\begin{eqnarray}
G _{st} ^{(1)} \left( t _{1} ; t _{2} \right) = 
\frac{ i  }{2 \pi  }
\frac{ 1 - \prod _{k = - \infty} ^{ \infty} S_{L} ^{*} \left( t_{1} + k {\cal T} _{0} \right)  S_{L}\left( t_{2} + k {\cal T} _{0} \right)  }{ t _{1} - t _{2} }  
\nonumber \\
\label{st-01-0} \\
= 
\frac{ \sin \left( \pi \frac{ t _{1} - t _{2} }{ {\cal T} _{0} } \right)  }{2\pi \left( t _{1} - t _{2} \right) }
\frac{ \sinh\left( \pi \frac{ 2  \Gamma _{\tau} }{ {\cal T} _{0} }  \right) }{ \sin \left( \pi \frac{ t _{1} + i \Gamma _{ \tau} }{ {\cal T} _{0} } \right) \sin \left( \pi \frac{ t _{2} - i \Gamma _{ \tau} }{ {\cal T} _{0} } \right) }
.
\nonumber 
\end{eqnarray}
\ \\
\noindent
The second line is given just for the further reference. 
However for subsequent calculations I need only the first line.

\subsection{Decomposition of the correlation function}

To bring Eq.~(\ref{st-01-0}) into the form of Eq.~(\ref{g1}) let us proceed as follows.
First, it is easy to verify the following identity, 

\begin{eqnarray}
1 - \prod _{k =1} ^{ \infty} x _{k} 
&=&  
(1 - x _{1})  + (1 - x _{2}) x _{1} + (1 - x _{3}) x _{1} x _{2} + \dots
\nonumber \\
\label{st-01-1} \\
&=&  
\sum\limits_{n=1}^{\infty} (1 - x _{n}) \prod _{k = 1} ^{ n-1} x _{k}. 
\nonumber 
\end{eqnarray}
\ \\
\noindent
Now one can see that the next identity is also correct,

\begin{eqnarray}
1 - \prod _{k =1} ^{ \infty} x _{k} 
=  
(1 - x _{1}) \prod _{k = 2} ^{ \infty} x _{k} 
+ (1 - x _{2}) \prod _{k = 3} ^{ \infty} x _{k} 
\nonumber \\
\label{st-01-2} \\
+ (1 - x _{3}) \prod _{k = 4} ^{ \infty} x _{k} + \dots 
=  \sum\limits_{n=1}^{\infty} (1 - x _{n}) \prod _{k = n+1} ^{ \infty} x _{k}. 
\nonumber 
\end{eqnarray}
\ \\
\noindent
The equation above is also valid if the lower limit is minus infinite, $k \to - \infty$. 

Then let us use Eq.~(\ref{st-01-2}) with an infinite lower limit and $x _{k} = S _{L} ^{*} (t _{1} + k {\cal T} _{0}) S _{L} (t _{2} + k {\cal T} _{0})  $, and rewrite Eq.~(\ref{st-01-0}) (its first line) as follows,

\begin{subequations}
\label{st02new}
\begin{eqnarray}
G _{st} ^{(1)} \left( t _{1} ; t _{2} \right) 
&=&
\sum\limits_{n=- \infty}^{ \infty} \Psi _{n} ^{*} (t _{1} ) \Psi _{n}  (t _{2} ) ,
\label{st-02-new} 
\end{eqnarray}
\ \\ \noindent
where the single-particle (envelope) wave functions,  

\begin{eqnarray}
\Psi _{n}  (t ) 
=
\Psi _{L} (t + n {\cal T} _{0}) \prod _{k = n+1} ^{ \infty} S _{L} (t + k {\cal T} _{0}) ,
\label{st-02-new-1} 
\end{eqnarray}
\ \\
\noindent
are orthonormal, 
\begin{eqnarray}
\int _{- \infty}^{ \infty } dt \Psi _{n} ^{*}(t) \Psi _{n^{\prime}} (t) = \delta _{n,n^{\prime}} .
\label{st-02-new-2} 
\end{eqnarray}
\end{subequations}
\ \\ \noindent
Here $\Psi _{L}(t) = \sqrt{ \Gamma _{ \tau} /  \pi } /\left( t - i \Gamma _{ \tau}  \right)$ is an envelope wave function of a single leviton in a one-dimensional channel.\cite{Keeling:2006hq,Grenier:2013gg} 
Note that the full leviton's wave function is $\psi_{L}(t) = \left( e^{-i \mu t/ \hbar}/ \sqrt{ v_{ \mu}} \right) \Psi _{L}(t)$. 
To get Eq.~(\ref{st-02-new}) we also need to use the following relation, $1 - S _{L} ^{*} (t _{1} + n {\cal T} _{0}) S _{L} (t _{2} + n {\cal T} _{0}) = - 2 \pi i (t _{1} - t _{2}) \Psi _{L} ^{*} (t _{1} + n {\cal T} _{0}) \Psi _{L} (t _{2} + n {\cal T} _{0})$.

Of course, the representation given in Eq.~(\ref{st-02-new}) is not unique, since the basis single-particle wave functions $\Psi_{n}$ given in Eq.~(\ref{st-02-new-1}) can be chosen in many different ways. 
The unitary transformations relating different bases can be time-dependent.

Note that the train of hole-like levitons is excited by the potential of an opposite sign, $V_{st}^{h}(t) = - V_{st}(t)$, where the superscript ``$h$'' denotes quantities related to holes. 
Formally it can be accounted by the following replacement, $  \Gamma _{\tau} \to -  \Gamma _{\tau}$, which also implies $S_{L}^{h}(t) = \left[ S_{L}(t) \right]^{*}$ and $\Psi_{L}^{h}(t) = \left [  \Psi_{L}(t)\right]^{*}$. 
As a result the analogue of expansion given in Eq.~(\ref{st-02-new}) acquires a minus sign, $G _{st} ^{(1),h} \left( t _{1} ; t _{2} \right) = - \sum_{n=- \infty}^{ \infty} \left [  \Psi _{n}^{h}(t _{1} ) \right]^{*}  \Psi _{n}^{h}  (t _{2} )$. 
The minus sign is a mere consequence of the fact that we consider an excess correlation function.

\subsection{Periodicity properties}

\begin{figure}[b]
\includegraphics[width=80mm]{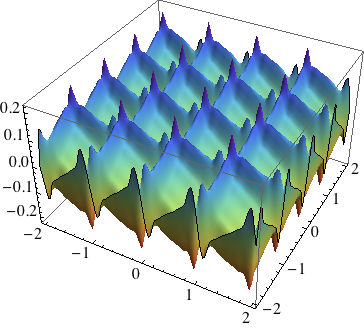}
\put(-242.0,100.0){\rotatebox{105}{\large ${\rm Re} g_{st}$}}
\put(-135.0,14.0){\large $t_{1}$}
\put(-25.0,40.0){\large $t_{2}$}
\caption{(Color online) The real part of the reduced first-order correlation function of the stream of levitons, $g_{st}\left( t_{1} ; t_{2} \right) = \left( t_{1} - t_{2} \right) G_{st}^{(1)}\left( t_{1}; t_{2} \right)$, is given as a function of its arguments $t_{1}$ and $t_{2}$ normalized to the period $ {\cal T} _{0}$. For $G_{st}^{(1)}$ see Eq.~(\ref{st-02-new}).  The width of a leviton is chosen to be $ 2 \Gamma_{ \tau}/ {\cal T} _{0} = 0.1$.}
\label{figG1lev}
\end{figure}

The basis wave functions $\Psi_{n}(t)$ in Eq.~(\ref{st-02-new-1}) are not periodic in time. 
Therefore, the correlation function $G_{st}^{(1)}\left( t_{1}; t_{2} \right)$, Eq.~(\ref{st-02-new}), is also not periodic in its arguments. 
However, the reduced correlation function, $g_{st}\left( t_{1} ; t_{2} \right) = \left( t_{1} - t_{2} \right) G_{st}^{(1)}\left( t_{1}; t_{2} \right)$, Eq.~(\ref{reduce}), is periodic in each of its arguments, see Fig.~\ref{figG1lev}. 

The wave functions $\Psi_{n}$ possess the following discrete time-translation symmetry,

\begin{eqnarray}
\Psi _{n}  (t + {\cal T} _{0} ) = \Psi _{n+1}  (t ), 
\label{ad-01}
\end{eqnarray}
\ \\
\noindent
which reflects the periodicity of a particle flux. 

The correlation function at coincident times is periodic,

\begin{eqnarray}
G ^{(1)}_{st}( t; t ) =  G ^{(1)}_{st}( t+ {\cal T}; t+{\cal T} ) 
= \sum\limits_{n=-\infty}^{\infty} \left | \Psi _{n}(t) \right | ^{2} .
\label{ad-02-1} 
\end{eqnarray}
\ \\
\noindent
Using Eq.~(\ref{ad-01}) one can then evaluate the mean number of electrons (levitons) per period,

\begin{eqnarray}
\left\langle N_{st} \right\rangle = 
\int _{0}^{ {\cal T} _{0} } dt G ^{(1)}_{st} \left( t; t \right) 
&=& \int _{0}^{ {\cal T} _{0} } dt \sum\limits_{n=-\infty}^{\infty}\left | \Psi _{n} (t) \right | ^{2}
\nonumber \\
\label{ad-03} \\
&=& \int _{ - \infty}^{ \infty } dt \left | \Psi _{L} (t) \right | ^{2} = 1,
\nonumber 
\end{eqnarray}
\ \\ \noindent
which is one, independently of the ratio $  \Gamma _{\tau}/ {\cal T} _{0}$, which characterises the degree of overlap between the levitons. 
  
Note that the periodicity of a flux  allows to express the correlation function of a single leviton, $G ^{(1)}_{L} \left( t_{1}; t_{2} \right) = \Psi _{L}^{*} (t_{1})\Psi _{L} (t_{2})$, in terms of an experimentally accessible correlation function of a stream, $G ^{(1)}_{st}\left( t_{1}; t_{2} \right)$. 
For this purpose we use Eqs.~(\ref{ad-01}), (\ref{ad-02-1}) and relate a discrete Fourier transformation, $G ^{(1)} _{st}(k)$, of a periodic function $G ^{(1)}_{st}\left( t; t \right)$  to a continuous Fourier transformation, $G ^{(1)}_{L} ( \omega)$, of a non-periodic function $G ^{(1)}_{L} \left( t; t \right)$, as follows,

\begin{eqnarray}
G ^{(1)} _{st}(k) 
&=& 
\int _{0}^{ {\cal T} _{0} } \frac{ dt }{ {\cal T} _{0} }  G ^{(1)}_{st}(t;t) e ^{i k \Omega t} 
\nonumber \\
&=& 
\int _{0}^{ {\cal T} _{0} } \frac{ dt }{ {\cal T} _{0} }  \sum\limits_{n=-\infty}^{\infty} \left | \Psi _{n}(t) \right | ^{2} e ^{i k \Omega t}
\nonumber \\
&=& 
\sum\limits_{n=-\infty}^{\infty} \int _{n {\cal T} _{0}}^{(n+1) {\cal T} _{0} } \frac{ dt^{\prime} }{ {\cal T} _{0} }   \left | \Psi _{L}(t^{\prime}) \right | ^{2} e ^{i k \Omega t^{\prime}} e ^{-i k \Omega n {\cal T} _{0}} 
\nonumber \\
&=& 
\int _{- \infty}^{ \infty } \frac{ dt^{\prime} }{ {\cal T} _{0} }   G ^{(1)}_{L} \left( t^{\prime}; t^{\prime} \right)  e ^{i k \Omega t^{\prime}} 
= \frac{1 }{ {\cal T} _{0} }  G ^{(1)}_{L} ( \omega = k \Omega),
\nonumber \\
\label{ad-05} 
\end{eqnarray}
\ \\
\noindent
where I made a substitution $t^{\prime} = t - n {\cal T} _{0}$ and took into account that $ {\cal T} _{0} \Omega = 2 \pi$, hence $\exp(-i k \Omega n {\cal T} _{0} ) = 1$. 
Note that a single-time Fourier transformation over $t = t_{1} = t_{2}$, $G ^{(1)} (k)$, is related to a two-time Fourier transformation introduced in Eq.~(\ref{pur-02-1}) as follows, $G ^{(1)} (k) = (1/ {\cal T} _{0}) \sum_{n=-\infty}^{\infty} G^{(1),in}\left( \epsilon_{n}, \epsilon_{n+k} \right)$.

Equation (\ref{ad-05}) defines $G ^{(1)}_{L} ( \omega)$ only on a discrete set of frequencies, $ \omega = k \Omega$. 
However, in the regime of adiabatic emission the Fourier coefficients $G ^{(1)}_{L} ( \omega)$ change smoothly on the scale of $ \Omega$.
This allows us to find $G ^{(1)}_{L} ( \omega)$ at other frequencies by interpolation and calculate,

\begin{eqnarray}
G ^{(1)}_{L} \left( t;t \right) 
&=& 
\int _{ - \infty}^{ \infty } \frac{d \omega  }{ \Omega }  G ^{(1)}_{st} \left( \frac{ \omega }{ \Omega } \right)  e ^{ - i \omega t} . 
\label{l-st} 
\end{eqnarray}
\ \\
\noindent
Finally using Eq.~(\ref{cf-05})  the full correlation function of a single leviton, $G ^{(1)}_{L}\left( t_{1}; t_{2} \right)$, can be reconstructed.

\subsection{The purity condition in time and energy representations}

From Eqs.~(\ref{st02new}) it immediately follows that Eq.~(\ref{conv-1}) (without a factor $v_{ \mu}$) holds, hence the state is pure. 
Therefore, the levitons are created in a deterministic fashion. 

The purity of a state can also be inferred from a correlation function in the  energy representation. 
Substituting Eq.~(\ref{st-02-new}) with wave functions from Eq.~(\ref{st-02-new-1}) into Eqs.~(\ref{pur-02-1}) and (\ref{pur-03-1}) we calculate. 

\begin{eqnarray}
G ^{(1),in}_{st} (\epsilon_{n} > 0; \epsilon_{\nu} >0) 
&=&
G ^{(1),out}_{st} (\epsilon_{n} > 0; \epsilon_{\nu} >0)
\nonumber \\
\label{inoutlev} \\
&=&
2 \sinh\left( \Omega \Gamma _{ \tau}  \right)
e ^{ - \Omega \Gamma _{ \tau}  \left(n + \nu + 1\right) } .
\nonumber 
\end{eqnarray}
\ \\
\noindent
Remember that $ \epsilon_{n} = \epsilon + n \hbar \Omega$ with the Floquet energy $0 < \epsilon < \hbar \Omega$. 
Therefore, $\epsilon_{n} > 0$ implies $n \geq 0$. 
For other energies the correlation function is zero, what is specific for purely electronic excitations.\cite{Bocquillon:2013fp}  
Note that $G ^{(1)} _{st}(k) = (1/ {\cal T} _{0}) \exp\left( - |k| \Omega  \Gamma _{\tau} \right)$. 

By applying the inverse Fourier transformation, Eq.~(\ref{pur-02}) or (\ref{pur-03}), to Eq.~(\ref{inoutlev}) we arrive at $G _{st} ^{(1)} \left( t _{1} ; t _{2} \right)$ in the form of the second line of Eq.~(\ref{st-01-0}), as it should be (see Appendix \ref{levFourier}). 
The two functions above, $G ^{(1),in}_{st}$ and $G ^{(1),out}_{st}$, do satisfy Eq.~(\ref{pur-06-new}) (without a factor $v_{ \mu}$) characteristic for pure states. 
Note that in the present case the sum over $m$ in Eq.~(\ref{pur-06-new}) runs from $m=0$. 

Thus, the purity of a state of the stream of levitons is demonstrated in both time and energy representations.

\subsection{Factorization in the time representation}
\label{p2}

The sum in Eq.~(\ref{st-02-new}) involves more than one term, therefore, in general  the state of the stream of levitons is a multi-particle state. 
As a result, all higher-order correlations functions are non zero. 
The multi-particle state is formed due to an overlap of the levitons excited during  different periods, which is the case when $  \Gamma _{\tau} \sim {\cal T} _{0}$. 

In contrast, if the levitons  do not overlap, which is the case when $  \Gamma _{\tau} \ll {\cal T} _{0}$ , then the state of a flux  is the direct product of states of individual levitons. 
This formally follows from Eq.~(\ref{st-02-new-1}), where $\lim_{\Gamma _{\tau}/ {\cal T} _{0} \to 0} S _{L} (t + k {\cal T} _{0}) = 1$ and, therefore, $\Psi _{n}  (t ) = \Psi _{L} (t + n {\cal T} _{0})$. 
The state with a wave function $\Psi _{n}(t)$ exists only at a period for which $t \approx - n {\cal T} _{0}$ and it vanishes at other periods.  

\begin{figure}[b]
\includegraphics[width=80mm]{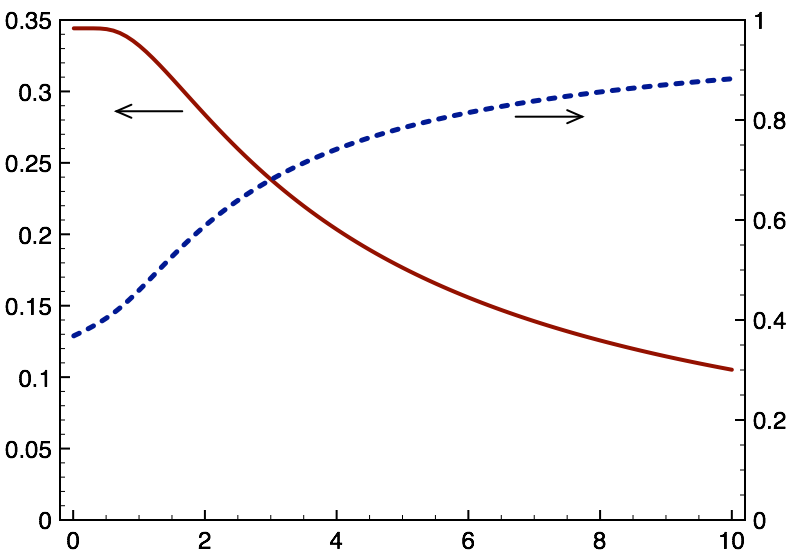}
\put(-240.0,140.0){$\left\langle N_{2} \right\rangle$}
\put(-8.0,140.0){$\mathbb{P}_{1}$}
\put(-130.0,-5.0){${\cal T} _{0}/  (2 \Gamma _{\tau} ) $}
\caption{(Color online) The mean number of pairs of levitons detected during a period,  $\left\langle N_{2} \right\rangle$, Eq.~(\ref{sps-08}), (a red solid line) and the probability to measure exactly one leviton during a period, $\mathbb{P}_{1}$, Eq.~(\ref{sps-04-1}), (a blue dashed line) are given as a function of the ratio of the time delay between two subsequent levitons, $ {\cal T} _{0}$, and the width of a leviton, $ 2 \Gamma_{ \tau}$. Notice the different axes used for $\left\langle N_{2} \right\rangle$ and $\mathbb{P}_{1}$, which are indicated by horizontal arrows. In the limit when  levitons do not overlap, ${\cal T} _{0}/(2\Gamma _{\tau}) \gg 1$, we have $\left\langle N_{2}\right\rangle\to~0$ and $\mathbb{P}_{1} \to 1$, which means that the multi-particle state is not formed.}
\label{figP2}
\end{figure}

In practice whether the states of electrons in the stream are single-particle states or not can be verified by examining whether the correlation function, $G^{(1)}\left( t_{1}; t_{2} \right)$ is factorized into two factors each of which depends only on one argument or not. 
If the first-order correlation function is factorizable then, as I already mentioned, all higher order correlation functions are zero, in particular, $G^{(2)} = 0$.  
Therefore, with increase of the ratio ${\cal T} _{0}/(2\Gamma _{\tau})$ the second order correlation function is expected to decrease. 
To confirm this expectation let us evaluate the mean number of pairs of levitons per period, $\left\langle N_{2} \right\rangle$, 

\begin{eqnarray}
\left\langle N_{2} \right\rangle  = 
\iint_{ 0 }^{ {\cal T} _{0} } dt dt ^{\prime}  G ^{(2)}_{st}  ( t ,t^{\prime} ;  t^{\prime}, t  )  ,
\label{sps-06}
\end{eqnarray}
\ \\
\noindent
which should decrease with decreasing the second-order correlation function $G ^{(2)}_{st}$. 
Representing $G ^{(2)}_{st}$ in terms of $G ^{(1)}_{st}$ according to Eq.~(\ref{cf-03}) and taking into account Eq.~(\ref{ad-03}), one can express $\left\langle N_{2} \right\rangle$ in terms of an experimentally accessibly first-order  correlation function, 

\begin{eqnarray}
\left\langle N_{2} \right\rangle
= 1 - \iint_{ 0 }^{ {\cal T} _{0} } dt dt ^{\prime} \left | G^{(1)}_{st} (t;t^{\prime}) \right |^{2} . 
\label{sps-08} 
\end{eqnarray}
\ \\
\noindent
where $G^{(1)}_{st}$ is given in Eq.~(\ref{st-01-0}) [or in Eq.~(\ref{st-02-new})].  
The dependence of $\left\langle N_{2} \right\rangle$ on the ratio ${\cal T} _{0}/  (2 \Gamma _{\tau} )$ is shown in Fig.~\ref{figP2} (a red solid line). 
The number of pairs of levitons, which can be detected during a period, gradually decreases with increasing the distance between the levitons, which demonstrates an evolution from a multi-particle state to a single-particle state. 
At the same time the probability to measure exactly one leviton during a period, $\mathbb{P}_{1}$, gradually increases up to one in the limit of ${\cal T} _{0}/  (2 \Gamma _{\tau} ) \to \infty$, see Fig.~\ref{figP2} (a blue dashed line). 
The details of calculations of $\mathbb{P}_{1}$ are given in Appendix \ref{p1}.

\subsection{Mean energy per leviton}

Another indication that the overlapping levitons do affect each other is an increase of the  mean  energy per leviton, $\left\langle \epsilon \right\rangle$, with increasing overlap. 
Using a single-particle distribution function for the stream of levitons, $f_{st}( \epsilon)   = G ^{(1),in}_{st} (\epsilon_{n}; \epsilon_{n}) \equiv f_{st}(n)$, Eq.~(\ref{inoutlev}), one can calculate,  

\begin{eqnarray}
\left\langle \epsilon \right\rangle 
&=& 
\frac{ \int _{}^{ } d \epsilon \epsilon f_{st}\left( \epsilon \right) }{ \int _{}^{ } d \epsilon f_{st} \left(  \epsilon \right) }
=
\hbar \Omega \sum\limits_{n=0}^{\infty} ( n + 0.5) f_{st}( n) 
\nonumber \\
\label{sps-09-1-1} \\
&=&  
\frac{ \hbar \Omega }{ 2 }
\coth\left( \Omega \Gamma _{\tau}   \right) 
=
\left\{  \begin{array}{ll}
\frac{ \hbar }{ 2  \Gamma _{\tau} }, &  \frac{\Gamma _{\tau} }{ {\cal T} _{0} } \ll 1 , \\
\ \\
\frac{ \hbar \Omega }{ 2 } , &   \frac{\Gamma _{\tau} }{ {\cal T} _{0} }  \gg 1 .
\end{array} \right. 
\nonumber 
\end{eqnarray}
\ \\
\noindent
The energy $ \left\langle \epsilon \right\rangle$ gradually increases with increasing overlap from the value $\hbar /(2\Gamma _{\tau})$\cite{Keeling:2006hq,Battista:2014di} characteristic for an isolated leviton to the value $ \hbar\Omega/2$ characteristic for a dc bias $ \hbar \Omega$. 
The latter is what $V_{st}(t)$, Eq.~(\ref{Vlev}), looks like at $  \Gamma _{\tau} \gg {\cal T} _{0}$.\cite{Hofer:2014va} 

\subsection{Accidental factorization in the energy representations}
\label{factor-e}

In contrast to a correlation function in the time representation, the correlation function of the stream of levitons in the energy representation, Eq.~(\ref{inoutlev}),  is factorized at any degree of overlap, $G_{st}^{(1),in} ( \epsilon_{n}; \epsilon_{ \nu}) =  \varphi_{st} (n)  \varphi_{st}( \nu)$, where $ \varphi_{st} (n) = e^{- n \Omega  \Gamma _{\tau} } \sqrt{2 \sinh\left( \Omega  \Gamma _{\tau} \right) e^{- \Omega  \Gamma _{\tau}}}$. 
Such a factorization results in vanishing of a two-particle distribution function (see Appendix \ref{2e}), $f_{st}( \epsilon_{n}, \epsilon_{ \nu}) = 0$. 
This result can be interpreted as a zero probability to measure two levitons with fixed energies, $ \epsilon_{n}$ and $ \epsilon_{ \nu}$.   
However this fact does not deny a possibility to measure two levitons during a  time period of duration $ {\cal T} _{0}$, see Fig.~\ref{figP2}.
The point is that a leviton has no a definite energy and its projection into the state with a definite energy can only be done via measurement over an infinite time.  
The possibility to measure two levitons during a finite time is also clearly demonstrated by the theory of waiting time distributions.\cite{Dasenbrook:2014tt} 

The factorization mentioned above is to some extend accidental and specific to a driving protocol given in Eq.~(\ref{Vlev}). 
For a general driving protocol the correlation function in the energy representation reveals the presence of multi-particle correlations. 
The energy representation is particularly useful to demonstrate the presence of correlations between electrons and holes created by an ac voltage. 
As an example the case of a harmonic driving voltage is shortly discussed in Appendix \ref{ac}.

\section{Conclusion}
\label{concl}

The excess first-order correlation function allows us to differentiate unambiguously whether the state of a periodic stream of electrons injected into a waveguide is a multi-particle state or can be represented as the product of single-particle states existing during each period separately. 
In the latter case the second- and all higher-order correlation functions, which all are expressed in terms of the first-order correlation function, vanish while in the former case they do not. 

Motivated by recent experiments\cite{Dubois:2013dv,Jullien:2014ii} I analysed a correlation function of a train of single electrons, levitons, excited by a periodic sequence of Lorentzian voltage pulses with a unit flux each. 
I decomposed this function into the sum of single-electron contributions. 
This allows us, first, to verify that the state of a stream is pure rather than mixed. 
Second, such a decomposition in the time representation allows us to show clearly that decreasing the width of a leviton makes the state of the electronic  stream evolving from a multi-particle state to a single-particle state. 
In the latter case the correlation function is factorizable while in the former case it is not.  
Unexpectedly I found that in the energy representation the correlation function is factorizable at any width of a leviton. 
This hinders a direct use of energy-resolved measurements to differentiate single- and multi-particle states in the case of a stream of levitons. 
The possible reason is that such measurements involve averaging over an infinite time interval, over which the single-particle wave functions are orthogonal due to the Fermi statistics irrespective of the degree of overlap.\cite{note1} 
Due to orthogonality even the overlapping particles contribute independently.

\acknowledgments
I thank G\'eraldine Haack for careful reading a manuscript and useful comments. 
Financial support from the Erasmus Mundus Program in Nanoscience and Nanotechnology is gratefully acknowledged. 
I appreciate the warm hospitality of the Institute for Materials Science of the 
TU Dresden where part of this work was done. 

\appendix

\begin{widetext}

\section{$(N+\dots)$th order correlation function for an $N$-electron state}
\label{proof2}

The determinant is calculated as follows,

\begin{eqnarray}
\det
\left(
\begin{array}{ccc}
 A _{ 11}   & \dots &   A _{ 1M}     \\
  \dots & \ddots  & \dots  \\
  A _{ M 1}  &  \dots & A _{ MM}     
\end{array}
\right) = 
\sum _{ j _{1},\cdots,j _{M} = 1 } ^{M}
\epsilon _{ j _{1} \dots j _{M} } A _{1 j _{1} }  \dots A _{M j _{M} } 
=
\sum _{ j _{1}, \cdots,j _{M} = 1 } ^{M}
\epsilon _{ j _{1}  \dots j _{M} } A _{j _{1} 1 } \dots A _{ j _{M} M } .
\label{det-1} 
\end{eqnarray}
\ \\ \noindent
Here $\epsilon _{ i _{1} i _{2} \dots i _{N} }$ is the Levi-Civita tensor with the following properties: (i) $\epsilon _{ 1 2 \dots N } = 1$, (ii) if any two indices are the same, then $\epsilon _{ \dots i _{1} \dots i _{1} \dots} = 0$, and, finally, (iii) when the two any indices are interchanged then the tensor changes a sign, $\epsilon _{ \dots i _{1} \dots i _{2} \dots} = - \epsilon _{ \dots i _{2} \dots i _{1} \dots}$. 
Using Eq.~(\ref{det-1}) in Eq.~(\ref{cf-03}) with $M \ne N$ we get,  
\begin{eqnarray}
\tilde G ^{(M)} _{N} (f _{1},\dots,f _{M};i_{M},\dots,i _{1}) 
&=& 
\sum _{ j _{1},\cdots,j _{M} = 1 } ^{M}
\epsilon _{ j _{1}  \dots j _{M} } 
\sum _{ k _{1},\dots,k _{M} = 1} ^{N} 
\psi _{ k _{1} } ^{*}(f _{1} ) \psi _{ k _{1} } (i _{ j _{1} } )
\dots
\psi _{ k _{M} } ^{*}(f _{M} ) \psi _{ k _{M} } (i _{ j _{M} } ) .
\label{gM-1} 
\end{eqnarray}
\ \\ \noindent
As it follows from Eq.~(\ref{g1}), there are $N$ different functions $\psi _{k}$, $k = 1,2,\dots,N$. 
Since $M > N$, then for any term in equation above among  $M$ factors $\psi _{ k _{\ell} } ^{*} \psi _{ k _{\ell} }$, $\ell = 1,2,\dots,M$, one can always find at least two factors, say for $\ell=\alpha$ and $\ell=\beta$, for which $k _{\alpha} = k _{\beta}$.   
Let us take a close look at the product of such two factors (we also keep the Levi-Civita tensor), 

\begin{eqnarray}
\epsilon _{\dots j _{\alpha}\dots j _{\beta} \dots } 
\psi _{ k _{\alpha} } ^{*}(f _{\alpha} ) \psi _{ k _{\alpha} } (i _{ j _{\alpha} } )\psi _{ k _{\beta} } ^{*}(f _{\beta} ) \psi _{ k _{\beta} } (i _{ j _{\beta} } ) 
&=& 
\epsilon _{\dots j _{\alpha}\dots j _{\beta} \dots } 
\psi _{ k _{\alpha} } ^{*}(f _{\alpha} ) \psi _{ k _{\alpha} } (i _{ j _{\alpha} } )\psi _{ k _{\alpha} } ^{*}(f _{\beta} ) \psi _{ k _{\alpha} } (i _{ j _{\beta} } )
\nonumber \\
&=&
- \epsilon _{\dots j _{\beta}\dots j _{\alpha} \dots } 
\psi _{ k _{\alpha} } ^{*}(f _{\alpha} ) \psi _{ k _{\alpha} } (i _{ j _{\beta} } )\psi _{ k _{\alpha} } ^{*}(f _{\beta} ) \psi _{ k _{\alpha} } (i _{ j _{\alpha} } ).
\label{kakb} 
\end{eqnarray}
\ \\ \noindent
Here we utilized the fact that $k _{\alpha} = k _{\beta}$ but in general $f _{ \alpha} \ne f _{ \beta}$ and $i _{ j _{\alpha} }   \ne i _{ j _{\beta} }$. 
Going from the second to the third line we made an interchange $j _{ \alpha} \leftrightarrow j _{ \beta}$ and took into account that the Levy-Civita tensor is antisymmetric, $\epsilon _{\dots j _{\alpha}\dots j _{\beta} \dots } = - \epsilon _{\dots j _{\beta}\dots j _{\alpha} \dots }$. 
The interchange $j _{ \alpha} \leftrightarrow j _{ \beta}$ does not change Eq.~(\ref{gM-1}), which involves sums over all $j_{ \alpha}$'s. 
On the other hand, according to Eq.~(\ref{kakb}), such an interchange should reverse a sign. 
These two consequences of the interchange $j _{ \alpha} \leftrightarrow j _{ \beta}$ are consistent if and only if the entire sum is zero.
Therefore, Eq.~(\ref{pure-NM}) is proven.

\section{The derivation of the purity condition in the energy representation, Eq.~(\ref{pur-06-new})}
\label{Fourie-pure}

Let us rewrite Eq.~(\ref{conv-1}) in terms of the Fourier coefficients of a correlation function. 
It is convenient to transform $\tilde G ^{(1)} _{} ( t _{1} ; t )$  according to Eq.~(\ref{pur-02}) and $\tilde G ^{(1)} _{} ( t ; t_{2} )$ according to Eq.~(\ref{pur-03}). 
As a result we arrive at the following,

\begin{eqnarray} 
\tilde G ^{(1)} _{} (t _{1} ;t _{2} )
&=&
e ^{ i \mu \frac{ \Delta t }{ \hbar }} 
v_{ \mu}
\int _{-\infty} ^{\infty} dt 
\int _{0}^{ \hbar \Omega } \frac{d \epsilon }{h }  e ^{ i \epsilon \frac{ t _{1} - t }{ \hbar }} 
\sum\limits_{ n=-\infty}^{\infty}
e ^{ i n \Omega \left( t _{1} - t \right) }
\sum\limits_{ m=-\infty}^{\infty}
e ^{ - i m \Omega   t } 
\label{pur-04-0} \\
&&\times
\int _{0}^{ \hbar \Omega } \frac{d \epsilon^{\prime} }{h }  e ^{ i \epsilon^{\prime} \frac{ t - t _{2} }{ \hbar }} 
\sum\limits_{ n ^{\prime}=-\infty}^{\infty}
e ^{ i  n ^{\prime} \Omega \left( t - t _{2} \right) }
\sum\limits_{ m ^{\prime}=-\infty}^{\infty}
e ^{ i m ^{\prime} \Omega   t } 
\tilde G ^{(1),out} (\epsilon_{n} ; \epsilon_{n+m}) 
\tilde G ^{(1),in} (\epsilon ^{\prime}_{n ^{\prime} + m ^{\prime}}; \epsilon ^{\prime}_{n ^{\prime}} )
\nonumber \\
&=& 
e ^{ i \mu \frac{ \Delta t }{ \hbar }} 
v_{ \mu}
\int _{0}^{ \hbar \Omega } \frac{d \epsilon }{h }  e ^{ i \epsilon \frac{ \Delta t }{ \hbar }} 
\sum\limits_{ n=-\infty}^{\infty}
\sum\limits_{ m=-\infty}^{\infty}
\sum\limits_{ m ^{\prime}=-\infty}^{\infty}
e ^{ i (n + m) \Omega t _{1} }
e ^{ - i (n + m ^{\prime} ) \Omega   t _{2} } 
\tilde G ^{(1),out} (\epsilon_{n+m};\epsilon_{n} ) 
\tilde G ^{(1),in} ( \epsilon_{n} ; \epsilon_{n + m^{\prime}} ),
\nonumber 
\end{eqnarray}
\ \\ \noindent
where $ \Delta t = t_{1} - t_{2}$. 
Note that the integral over $t$ produces 
$h \delta \left( - \epsilon - (n+m)\hbar \Omega + \epsilon^{\prime} + (n^{\prime} + m^{\prime} )\hbar \Omega \right)$. 
Since both Floquet energies $ \epsilon ^{\prime}$ and $ \epsilon$ belong to the same interval of length $ \hbar \Omega$, this delta function implies $\epsilon^{\prime} = \epsilon$ and $n+m= n^{\prime} + m^{\prime}$.   
In the last line of Eq.~(\ref{pur-04-0}) we additionally changed $n+ m \to n$,  $ m \to - m$, and $ m^{\prime} \to - m^{\prime}$. 

To proceed let us perform the Fourier transformation on the left hand side either for in- or for out- representation using Eqs.~(\ref{pur-03-1}) or (\ref{pur-02-1}), respectively. 
First we use Eq.~(\ref{pur-03-1}), put $t _{2} = t _{1} - \Delta t $, and find

\begin{eqnarray}
\tilde G ^{(1),in} (\epsilon _{ \ell+\nu}  ; \epsilon _{ \ell} ) 
&=&
v_{ \mu}
\int _{ - \infty}^{ \infty } d \Delta t   e ^{- i (\epsilon _{ \ell} - \mu )   \frac{ \Delta t }{ \hbar }} 
\int _{ 0 }^{ {\cal T} _{0} } \frac{ d t _{1} }{ {\cal T} _{0} }
e ^{  -i \nu \Omega   t _{1}} 
\int _{0}^{ \hbar \Omega } \frac{d \epsilon^{\prime} }{h }  e ^{ i \epsilon^{\prime} \frac{ \Delta t }{ \hbar }} 
\sum\limits_{ n=-\infty}^{\infty}
\sum\limits_{ m=-\infty}^{\infty}
\sum\limits_{ m ^{\prime}=-\infty}^{\infty}
e ^{ i (n + m) \Omega t _{1} }
e ^{ - i (n + m ^{\prime} ) \Omega   (t _{1} - \Delta t) } 
\nonumber \\
&&\times
\tilde G ^{(1),out} (\epsilon^{\prime} _{n + m} ; \epsilon^{\prime} _{n} ) 
\tilde G ^{(1),in} (\epsilon^{\prime} _{  n } ; \epsilon^{\prime} _{ n + m^{\prime} } )
= 
v_{ \mu}
\sum\limits_{ m^{\prime}=-\infty}^{\infty}
\tilde G ^{(1),out} (\epsilon _{ \ell + \nu} ; \epsilon _{ \ell - m^{\prime}} ) 
\tilde G ^{(1),in} ( \epsilon _{ \ell - m^{\prime}} ; \epsilon _{ \ell }  ) .
\label{pur-05} 
\end{eqnarray}
\ \\ \noindent
where $ \ell$, $ \nu$ are integers. 
Next we use Eq.~(\ref{pur-02-1}), put $t _{1} = \Delta t + t _{2}$, and arrive at the following,

\begin{eqnarray}
\tilde G ^{(1),out} (\epsilon _{ \ell} ; \epsilon _{ \ell+\nu}) 
&=&
v_{ \mu}
\int _{ - \infty}^{ \infty } d \Delta t   e ^{- i (\epsilon _{ \ell} - \mu )  \frac{ \Delta t }{ \hbar }} 
\int _{ 0 }^{ {\cal T} _{0} } \frac{ d t _{2} }{ {\cal T} _{0} }
e ^{  i \nu \Omega   t _{2}} 
\int _{0}^{ \hbar \Omega } \frac{d \epsilon^{\prime} }{h }  e ^{ i \epsilon^{\prime} \frac{ \Delta t  }{ \hbar }} 
\sum\limits_{ n=-\infty}^{\infty}
\sum\limits_{ m=-\infty}^{\infty}
\sum\limits_{ m ^{\prime}=-\infty}^{\infty}
e ^{ i (n + m) \Omega ( \Delta t + t _{2}) }
e ^{ - i (n + m ^{\prime} ) \Omega   t _{2} } 
\nonumber \\
&&\times
\tilde G ^{(1),out} (\epsilon^{\prime} _{n + m} ; \epsilon^{\prime} _{n} ) 
\tilde G ^{(1),in} (\epsilon^{\prime} _{  n } ; \epsilon^{\prime} _{ n + m^{\prime} } )
=
v_{ \mu}
\sum\limits_{ m=-\infty}^{\infty}
\tilde G ^{(1),out} (\epsilon _{ \ell} ; \epsilon _{ \ell - m} ) 
\tilde G ^{(1),in} ( \epsilon _{ \ell - m} ; \epsilon _{ \ell + \nu}  ) ,
\label{pur-04} 
\end{eqnarray}
\ \\ \noindent
So from Eqs.~(\ref{pur-05}) and (\ref{pur-04}) we get,

\begin{eqnarray}
\tilde G ^{(1),in} (\epsilon_{n} ; \epsilon_{\nu}) 
&=&
v_{ \mu}
\sum\limits_{ m=-\infty}^{\infty}
\tilde G ^{(1),out} (\epsilon_{n} ; \epsilon_{m} ) 
\tilde G ^{(1),in} ( \epsilon_{m}; \epsilon_{\nu}  ) ,
\nonumber \\
\tilde G ^{(1),out} (\epsilon_{n} ; \epsilon_{\nu} ) 
&=& 
v_{ \mu}
\sum\limits_{ m=-\infty}^{\infty}
\tilde G ^{(1),out} ( \epsilon _{n} ; \epsilon _{m} ) 
\tilde G ^{(1),in} ( \epsilon _{m} ; \epsilon _{\nu}  ) .
\label{pur-06} 
\end{eqnarray}
\ \\ \noindent
Since the right hand sides of these equations are the same, the left hand sides are also the same. 
Therefore, we arrive at Eq.~(\ref{pur-06-new}).

\section{The inverse Fourier transformation of the correlation function of the stream of levitons}
\label{levFourier}

Let us apply the inverse Fourier transformation, say, Eq.~(\ref{pur-02}), to Eq.~(\ref{inoutlev}), where only $n \ge 0$ and $\nu= m + n \ge 0$ do contribute. 
As a result we obtain the following equation, 

\begin{eqnarray}
G _{st} ^{(1)} \left( t_{1}; t _{2}  \right)  
&=&
2 \sinh\left( \Omega \Gamma _{ \tau}  \right)
\int _{ 0 }^{ \hbar \Omega } \frac{ d \epsilon }{h } e ^{ i \epsilon \frac{ \Delta t }{ \hbar } }
\sum\limits_{n= 0}^{\infty} e ^{ i n \Omega \Delta t  }
\sum\limits_{ \nu =0 }^{\infty} e ^{ - i ( \nu - n) \Omega t _{2} }
e ^{ - \Omega \Gamma _{ \tau}    \left( n + \nu + 1  \right)  }
\label{inv-3} \\
&=& 
\frac{ \sinh\left( \Omega \Gamma _{ \tau}  \right) \left\{ e^{i \Omega \Delta t} -1 \right\} }{ i \pi \Delta t }
\frac{ 1 }{ 1 - e^{ \Omega [i t_{1} -  \Gamma _{\tau}]} }
\frac{ e^{ -  \Gamma _{\tau} \Omega} }{ 1 - e^{ -\Omega [i t_{2} +  \Gamma _{\tau}]} }
= 
\frac{ \sin \left( \Omega  \frac{ \Delta t }{ 2 } \right)  }{ 2\pi  \Delta t }
\frac{ \sinh\left(\Omega \Gamma _{ \tau}   \right) }{ \sin \left( \Omega  \frac{ t _{1} + i \Gamma _{ \tau} }{ 2 } \right) \sin \left( \Omega  \frac{ t _{2} - i \Gamma _{ \tau} }{ 2 } \right) } \, ,
\nonumber 
\end{eqnarray}
\ \\ \noindent
which coincides with the second line of Eq.~(\ref{st-01-0}). 

\section{Probability to measure exactly one leviton per period}
\label{p1}

The probability, $p_{n}$, to detect a leviton with a wave function $\Psi_{n}$, Eq.~(\ref{st-02-new-1}), during the time interval $\left( 0 ; {\cal T} _{0} \right)$ is the following,  

\begin{eqnarray}
p _{n} 
&=&   
\int _{ 0  }^{ {\cal T} _{0} } dt  \left | \Psi _{n} (t) \right | ^{2}
= 
\int _{  n {\cal T} _{0}}^{  (n+1) {\cal T} _{0} } dt  \tilde G^{(1)}_{L}\left( t;t \right)
=
\frac{ \arctan\left( \frac{ (n+1) {\cal T} _{0} }{ \Gamma _{\tau}} \right)  - \arctan\left( \frac{ n{\cal T} _{0}  }{   \Gamma _{\tau} } \right)  }{ \pi }  .
\label{sps-02} 
\end{eqnarray}
\ \\ \noindent
The sum of all $p_{n}$'s gives the mean number of levitons detected during a period, $\sum_{- \infty}^{ \infty}p_{n} = \left\langle N_{st} \right\rangle =~1$, which agrees with Eq.~(\ref{ad-03}).
However the actual number of detected levitons varies from period to period. 
In particular, the probability to measure exactly one leviton per period  irrespective of its state is given by the following equation,  

\begin{eqnarray}
\mathbb{P}_{1} &=& \sum\limits_{n=-\infty}^{\infty} p_{n} \prod_{k\ne n, k=- \infty}^{ \infty} (1 - p_{k}) = \mathbb{P}_{0} \sum\limits_{n=-\infty}^{\infty} \frac{ p_{n} }{ 1 - p_{n} } , 
\label{sps-04-1} 
\end{eqnarray}
\ \\ \noindent
where $\mathbb{P}_{0} = \prod_{k=- \infty}^{ \infty} (1 - p_{k})$ is a probability to do not measure any leviton.
In the equation above the probability $p_{n}$ to detect a leviton with a wave function $\Psi_{n}$ is weighted by the product of probabilities, $(1 - p_{k})$, to do not detect a leviton in any other state, $k \ne n$.
Note that $\mathbb{P}_{1} = \left\langle N_{st} \right\rangle = 1$ if and only if all $p_{k} = 0$ but some $p_{n} = 1$. 
This is the case in the absence of overlaps, $ {\cal T} _{0}/ (2 \Gamma _{\tau}) \to \infty$, which is clearly seen from Fig.~\ref{figP2} (a blue dashed line). 
In this limit the number of particles detected during a given period does not fluctuate.

\section{Two-particle distribution function}
\label{2e}

The two-particle distribution function is defined as follows, \cite{Moskalets:2014ea} 

\begin{eqnarray}
f( \epsilon_{n}, \epsilon_{n+ m}) 
&=&  
\int _{0}^{ {\cal T} _{0} } \frac{dt }{ {\cal T} _{0} }\int _{0}^{ {\cal T} _{0} } \frac{dt^{\prime} }{ {\cal T} _{0} } \int _{ -\infty}^{ \infty } d \tau e^{ - i  \epsilon_{n} \frac{ \tau  }{ \hbar} } \int _{ -\infty}^{ \infty } d \tau^{\prime} e^{ - i \epsilon_{n+m} \frac{ \tau^{\prime}   }{  \hbar} } G^{(2)}( \tau + t,\tau^{\prime} + t^{\prime};t^{\prime},t) 
\nonumber \\
&=& 
f( \epsilon_{n}) f( \epsilon_{n+m}) + \delta^{2} f( \epsilon_{n}, \epsilon_{n+m}), 
\label{sps-09-2} 
\end{eqnarray}
\ \\ \noindent
where $f( \epsilon_{n}) = G^{(1),in}\left( \epsilon_{n}; \epsilon_{n} \right)= G^{(1),out}\left( \epsilon_{n}; \epsilon_{n} \right)$ is a single-particle distribution function. 
The irreducible part reads, 

\begin{eqnarray}
 \delta^{2} f( \epsilon_{n}, \epsilon_{n+m}) = (-1)  \int _{0}^{ {\cal T} _{0} } \frac{dt }{ {\cal T} _{0} }\int _{0}^{ {\cal T} _{0} } \frac{dt^{\prime} }{ {\cal T} _{0} } \int _{ -\infty}^{ \infty } d \tau e^{ - i \epsilon_{n} \frac{ \tau   }{ \hbar} }  \int _{ -\infty}^{ \infty } d \tau^{\prime} e^{ - i  \epsilon_{n+m}   \frac{ \tau^{\prime} }{ \hbar} } G^{(1)}( \tau + t; t^{\prime}) G^{(1)}( \tau^{\prime} + t^{\prime}; t). 
\label{sps-09-3} 
\end{eqnarray}
\ \\ \noindent
After the following shifts, $ \tau \to \tau - t + t^{\prime}$ and $ \tau^{\prime} \to \tau^{\prime} - t^{\prime} + t$, the equation above becomes, 

\begin{eqnarray}
 \delta^{2} f( \epsilon_{n}, \epsilon_{n+m}) 
 &=& 
 (-1)  
 \int _{0}^{ {\cal T} _{0} } \frac{dt^{\prime} }{ {\cal T} _{0} } e^{i m \Omega t^{\prime} }   \int _{ -\infty}^{ \infty } d \tau e^{ - i  \epsilon_{n}  \frac{ \tau  }{ \hbar} }  
G^{(1)}( \tau + t^{\prime}; t^{\prime}) 
 \int _{0}^{ {\cal T} _{0} } \frac{dt }{ {\cal T} _{0} } e^{- i m \Omega t}  \int _{ -\infty}^{ \infty } d \tau^{\prime} e^{ - i  \epsilon_{n+m}  \frac{ \tau^{\prime}  }{ \hbar} }  G^{(1)}( \tau^{\prime} + t; t)
\nonumber \\
&=& 
(-1)  G^{(1),out} ( \epsilon_{n}; \epsilon_{n+m})  G^{(1),out} ( \epsilon_{n+m}; \epsilon_{n})  ,
\label{sps-09-4} 
\end{eqnarray}
\ \\ \noindent
where we used Eq.~(\ref{pur-02-1}) written in terms of $G^{{(1)}}$ not $\tilde G^{{(1)}}$. 
Substituting the equation above into Eqs.~(\ref{sps-09-2}) we arrive at the following equation, 

\begin{eqnarray}
f( \epsilon_{n}, \epsilon_{ \nu}) = G ^{(1),out} (\epsilon_{n}; \epsilon_{n})  G ^{(1),out} (\epsilon_{ \nu}; \epsilon_{ \nu}) - G ^{(1),out} (\epsilon_{n}; \epsilon_{ \nu})  G ^{(1),out} (\epsilon_{ \nu}; \epsilon_{ n}), 
\label{fe-cor-in} 
\end{eqnarray}
\ \\ \noindent
which is used in Sec.~\ref{factor-e}. 

\end{widetext}

\section{First-order correlation function of an electron-hole flux  generated by an AC voltage}
\label{ac}

In contrast to a periodic sequence of Lorentzian voltage pulses, Eq.~(\ref{Vlev}), the general ac voltage excites both electrons and holes.   
As an example let us consider a harmonic voltage, $V_{ac}(t) = V \cos\left( \Omega t \right)$, applied across a single-channel ballistic electronic waveguide connected to two metallic contacts.
The corresponding scattering amplitude reads (up to an unimportant constant phase factor), $S _{ac}(t) = \exp\left\{ - i \frac{ eV }{ \hbar \Omega } \sin( \Omega t)\right\}$.
Without the loss of generality one can put $eV > 0$ and use the following expansion, 
\begin{eqnarray}
e ^{ i \frac{ eV }{ \hbar \Omega }  sin( \Omega t) } = \sum\limits_{n=-\infty}^{\infty} J _{n}\left( \frac{ eV }{ \hbar \Omega }  \right) e ^{ i n \Omega t} , 
\label{acnew01}
\end{eqnarray}
where $J_{n}$ is the Bessel function of the first kind and of the integer order $n$. 
Then we represent the correlation function, see Eq.~(\ref{cf-01}), as follows,

\begin{eqnarray}
G _{ac} ^{(1)}(t _{1} ; t _{2}) 
=
\frac{ i  }{2 \pi  }
\frac{ 1 - e ^{  i \frac{ eV }{ \hbar \Omega } \left\{  \sin( \Omega t_{1}) -  \sin( \Omega t_{2})   \right\}   } }{ t _{1} - t _{2} }
\nonumber \\
=
\sum\limits_{n, m=-\infty}^{\infty} 
J _{n} \left( \frac{ eV }{ \hbar \Omega } \right) J _{n + m} \left( \frac{ eV }{ \hbar \Omega } \right) e ^{- i \Omega m t _{2}}   
\int _{0}^{ n \hbar \Omega } \frac{ d \epsilon }{h } e ^{i \frac{ \epsilon }{ \hbar } \Delta t }  ,
\nonumber \\
=
\sum\limits_{n, m=-\infty}^{\infty} 
J _{n} \left( \frac{ eV }{ \hbar \Omega } \right) J _{n + m} \left( \frac{ eV }{ \hbar \Omega } \right) e ^{ i \Omega m t _{1}}   
\int _{0}^{ n \hbar \Omega } \frac{ d \epsilon }{h } e ^{i \frac{ \epsilon }{ \hbar } \Delta t }  ,
\nonumber \\
\label{ac-03} 
\end{eqnarray}
\ \\ \noindent
where we also used the following identity, 

\begin{eqnarray}
\sum _{n=-\infty}^{\infty}  J _{n+k}  \left( \frac{ eV }{ \hbar \Omega } \right)J _{n + m}  \left( \frac{ eV }{ \hbar \Omega } \right) = \delta _{k,m} .
\label{bess4}
\end{eqnarray}
Then we split the sum over $n$ in Eq.~(\ref{ac-03}) into the two parts, for $n>0$ and $n<0$, and arrive at the following,

\begin{eqnarray}
G _{ac} ^{(1)}(t _{1} ; t _{2})  
&=& 
\sum\limits_{m=-\infty}^{\infty} \sum\limits_{n=0}^{\infty}  \sum\limits_{k=n+1}^{\infty} 
\left\{ A_{m,n,k} - (-1)^{m} A_{m,n,k}^{*} \right\}, 
\nonumber \\
A_{m,n,k}  
&=& 
G^{(1)}_{ n}\left( \Delta t \right) 
J _{k} \left( \frac{ eV }{ \hbar \Omega } \right) J _{k + m} \left( \frac{ eV }{ \hbar \Omega } \right) 
 e ^{- i \Omega m t _{2}} 
\nonumber \\
&=&
G^{(1)}_{ n}\left( \Delta t \right)
J _{k} \left( \frac{ eV }{ \hbar \Omega } \right) J _{k + m} \left( \frac{ eV }{ \hbar \Omega } \right)  e ^{ i \Omega m t _{1}}   , 
\label{ehAC} 
\end{eqnarray}
with 
\begin{eqnarray}
G^{(1)}_{ n}\left( \Delta t \right) = \int _{ n \hbar \Omega }^{ (n+1)\hbar \Omega} \frac{ d \epsilon^{\prime} }{ h } e ^{ i \epsilon^{\prime} \frac{ \Delta t }{ \hbar }} .
\label{ac-basis}
\end{eqnarray}
\ \\
\noindent
The correlation functions $G^{(1)}_{ n}$ for different $n$ are mutually orthogonal and do satisfy the purity condition, Eq.~(\ref{conv-1}). 
Therefore, the corresponding states  constitute a convenient basis for representing the state generated by a harmonic voltage.\cite{Dubois:2013fs}  

As I already mentioned, electron and hole contributions to the correlation function have different signs, which hinders a direct use of Eq.~(\ref{conv-1}) for the state purity verification.  
To assess electronic and hole contributions separately it is useful to go over from a time domain to an energy domain.
After performing the Fourier transformation according to Eqs.~(\ref{pur-03-1}), (\ref{pur-02-1}) we have, 

\begin{eqnarray}
G ^{(1),in}_{ac} (\epsilon_{n}; \epsilon_{ \nu}) = G ^{(1),out}_{ac} (\epsilon_{ n } ; \epsilon_{ \nu }) =
\nonumber \\
\nonumber \\
= 
\left\{
   \begin{array}{ll} 
      \sum\limits_{k= 1}^{\infty} J _{k + n}\left( \frac{ eV }{ \hbar \Omega } \right) J _{k + \nu }\left( \frac{ eV }{ \hbar \Omega } \right),  & n \ge 0, \\
      \\
      -\sum\limits_{k= 0}^{-\infty} J _{k + n}\left( \frac{ eV }{ \hbar \Omega } \right) J _{k + \nu }\left( \frac{ eV }{ \hbar \Omega } \right),  & n \le -1. 
   \end{array}
\right.
\label{AC-e-in} 
\end{eqnarray}
\ \\ \noindent
Here both $n$ and $ \nu$ are integers. 
Remember that $ \epsilon_{n} = \epsilon + n \hbar \Omega$ and the Floquet energy $ \epsilon$ is chosen to be positive, $0 < \epsilon < \hbar \Omega$.

If in the equation above both energies belong to either electron, $ \epsilon_{n}, \epsilon_{ \nu} > 0$, or  hole, $ \epsilon_{n}, \epsilon_{ \nu} < 0$, sectors the equation (\ref{pur-06-new}) holds, what tells us that the emitted state is pure. 
Note that the sum on the right hand side (RHS) of the second line of Eq.~(\ref{pur-06-new}) comprises both purely electron (hole) and electron-hole contributions.
Therefore, both an electron coherence and an electron-hole coherence are equally important to maintain the entire state pure. 
A non-zero value of $G ^{(1),in}_{ac}$ in the electron-hole sector, $ \epsilon_{n} \ge 0, \epsilon_{ \nu} \le 0$ or $ \epsilon_{n} < 0, \epsilon_{ \nu} > 0$, indicates the presence of electron-hole correlations.\cite{Grenier:2011js,Grenier:2011dv,Bocquillon:2013fp} 
Note that in the electron-hole sector the sum on the RHS of the second line of Eq.~(\ref{pur-06-new}) is zero due to cancellation of electron and hole contributions, which is specific for states with equal number of electrons and holes.


\begin{thebibliography}{11}
\bibitem{note1}
Due to the same reason the zero-frequency electrical noise (the shot noise) is not sensitive to the degree of overlap of levitons.
\end{thebibliography}
\end{document}